\documentclass[12pt]{iopart}
\usepackage{amsfonts,amssymb,amsbsy,amsthm,graphicx}

\usepackage{braket}
\usepackage{dcolumn}
\usepackage{bm}

\newcommand{\kms}[1]{\mbox{\scriptsize #1}}

\begin{document}

\newtheorem{theo}{Theorem} \newtheorem{lemma}{Lemma}
\newcommand{\mss}[1]{\mbox{\scriptsize #1}}

\title{Optimal measurement-based feedback control for a single qubit: a candidate protocol}

\author{Ashkan Balouchi$^1$ and Kurt Jacobs$^{1, 2}$} 
\address{ 
$^1$Physics Department and Hearne Institute for Theoretical Physics, Louisiana State University, Baton Rouge, LA 70803, USA \\
$^2$Department of Physics, University of Massachusetts at Boston, Boston, MA 02125, USA 
}

\ead{kurt.jacobs@umb.edu} 

\begin{abstract}
Feedback control of quantum systems via continuous measurement involves complex nonlinear dynamics. Except in very special cases, even for a single qubit optimal feedback protocols are unknown. Not even do intuitive candidates exist for choosing the measurement basis, which is the primary non-trivial ingredient in the feedback control of a qubit. Here we present a series of arguments that suggest a particular form for the optimal protocol for a broad class of noise sources in the regime of good control. This regime is defined as that in which the control is strong enough to keep the system close to the desired state. With the assumption of this form the remaining parameters can be determined via a numerical search.  The result is a non-trivial feedback protocol valid for all feedback strengths in the regime of good control. We conjecture that this protocol is optimal to leading order in the small parameters that define this regime. The protocol can be described relatively simply, and as a notable feature contains a discontinuity as a function of the feedback strength. 
\end{abstract}

\maketitle 

\section{Introduction}

Tremendous experimental progress has been made in the last few years in the real-time measurement of mesoscopic systems. The development of parametric amplifiers with very low noise~\cite{Bergeal10, Hatridge11, Bergeal12} has allowed single qubits to be observed in real-time~\cite{Yu08, Vijay11}, culminating recently in the first realizations of continuous-time feedback control of a single mesoscopic qubit~\cite{Vijay12, Brakhane12}. Considerable experimental progress is also being made in the feedback control of microscopic systems~\cite{Gillett10, Koch10, Sayrin11, Brakhane12}. 

It is timely therefore to reflect on the state of the theory of continuous-time measurement-based control of simple quantum systems~\cite{HFHofmannOptExp:1998, HFHofmannPRA:1998, JWangPRA:2001, Wiseman02, RRuskovPRB:2002, rapidP, Ralph04, ANKorotkovMicroJ:2005, ANKorotkovPRB:2005, vanHandel05, QZhangPRB:2005, ANJordanPRB:2006, JSJinPRB:2006, Griffith06x, vanHandel07, SCHouPRA:2010, GKieblichPRL:2011}. While progress has been made in understanding the dynamics induced by continuous measurements, and its implications for feedback control~\cite{rapidP, Combes06, Griffith06x, Ashhab10, Jacobs2010, Combes10, Combes11}, except in certain special cases~\cite{Shabani08, Jacobs08b, Combes12x} it is still unknown how to use feedback to best control a single qubit. A feedback protocol involves continuously measuring an observable of the qubit, and modifying the Hamiltonian of the qubit with time. Specifically, the ``protocol'' is the rule by which we choose the observable to measure,and the Hamiltonian, at each time as a function of the state of the system (the density matrix). The problem of finding a superior feedback protocol is not in choosing the Hamiltonian at each time: although as yet unproven, so long as there is no restriction on what observable can be measured, and the noise is not unusually asymmetric, it is obvious that the optimal Hamiltonian is the one that moves the state closest to the desired state at each timestep. When the only restriction on the Hamiltonian is the speed that it can rotate on the Bloch sphere, this is achieved by following a geodesic on this sphere. It is the problem of how to chose the observable to measure as a function of the current state of the qubit that has no obvious answer. 

By contrast to open-loop control, the problem of finding optimal feedback protocols is very difficult to solve numerically. To do so one must solve the Hamilton-Jacobi-Bellman equation~\cite{Jacobs08b, Jacobs14}, which involves optimizing at the final infinitesimal time-step, and then stepping back, one time-step at a time, optimizing at each time until we reach the initial time. The size of the search space is the number of timesteps used to discretize time, multiplied by the number of grid points used to discretize the space of control options for each possible state at each time. The density matrix for a single qubit has three real parameters, and the measurement has three real parameters, so the space of control options at each time-step is 6-dimensional. The resulting numerical optimization problem is daunting. 

Here we address the problem of finding a feedback protocol that is optimal in the steady-state. We do not concern ourselves especially with the optimality of the protocol in the initial, transient regime. We also specialize our analysis in two ways, both of which will simplify the problem to some extent. The first is that we restrict our attention to the regime of good control. This regime is defined as that in which the control forces are sufficient to keep the system close to the desired state, $|\psi\rangle$, during the relevant time interval~\cite{Jacobs07c, Li09}. More precisely, ``close'' means that the probability of finding the system in a state orthogonal to $|\psi\rangle$ is much less than unity. Since we are interested here in optimality only in the steady-state, we also require that the protocol is in the regime of good control only in the steady-state. The initial state of the system may be anything.  

Our second specialization is that we restrict our analysis to Markovian noise processes that are symmetric about the $z$-axis of the qubit. This is a broad class of processes. In fact, all the commonly considered noise processes have this symmetry: dephasing, decay, thermalization, and depolarizing~\cite{Jacobs14}. This symmetry provides an initial simplification, especially when the state in which we wish to place the qubit --- the target state --- is also symmetric about the $z$-axis (is an eigenstate of $z$). If the target state is $|0\rangle$ or $|1\rangle$, then one parameter is eliminated from the density matrix, reducing the number of parameters to five. We will find that in the regime of good control, the above symmetry allows a further reduction to four parameters.  

Our method is to show that by focussing on the regime of good control, and analyzing the dynamics under measurement, one can make a number of well-motivated conjectures about the form the optimal protocol should take. Assuming these conjectures to be true leaves only a single parameter of the measurement basis to be chosen as a function of a single parameter of the state. The resulting optimization problem can be tackled readily with numerical simulations. Performing this optimization we find that the numerical results are sufficiently simple that the resulting protocol, can be specified analytically. This is the first example of a nontrivial protocol for the general control of a qubit under finite control speed, and we conjecture that it is optimal in the regime of good control. 

In Section~\ref{setup} we present the description of the qubit, the Hamiltonian, and the continuous measurement, as well as defining the regime of good control in terms of our description. In Section~\ref{args} we present the arguments that suggest a form for the optimal feedback protocol. In Section~\ref{nums} we perform the numerical optimization, and present our candidate optimal feedback protocol. In Section~\ref{disc} we finish by providing some intuitive arguments as to why our candidate protocol has the form it does, in terms of known properties of continuous measurements. 


\section{Parametrizing the qubit and the measurement}
\label{setup}

As discussed in the introduction, feedback control involves making a continuous measurement of an observable of the qubit, and modifying the Hamiltonian of the qubit with time. Since all observables for a single qubit can be written as a sum of the three Pauli operators, we can denote the measured observable by 
\begin{equation}
   \sigma_{\mathbf{m}} = \mathbf{m}\cdot \boldsymbol{\sigma}, 
\end{equation}
where $\mathbf{m}$ is a real three-dimensional unit vector, and $\boldsymbol{\sigma} = (\sigma_x, \sigma_y, \sigma_z)$ is the vector of Pauli matrices. The density matrix of the qubit can be written in terms of the three-dimensional Bloch vector, $\mathbf{a} = (a_x,a_y,a_z)$ as 
\begin{equation}
  \rho = \frac{1}{2}(I + \mathbf{a}\cdot\boldsymbol{\sigma}) . 
\end{equation}
Since our control problem is symmetry about the $z$-axis the $x$ and $y$ directions are equivalent, and we can eliminate the $y$ direction. We can then write the density matrix in terms of the length of the Bloch vector, $a \equiv |\mathbf{a}|$, and the angle between it and the ground state, which we will denote by $\theta$. With these definitions the density matrix is given by 
\begin{equation}
  \rho = \frac{1}{2}[I + a (\sin\theta \sigma_x - \cos\theta \sigma_{z})] . 
  \label{rhoxz}
\end{equation}
The evolution of the density matrix under the measurement is given by the stochastic master equation 
(SME)~\cite{Jacobs2006, Brun2002},  
\begin{eqnarray}
  d\rho &= & - \frac{i}{\hbar}[H(t),\rho]dt  -k [\sigma_{\mathbf{m}},[\sigma_{\mathbf{m}},\rho]]dt \nonumber \\ 
  & & + \; \sqrt{2k}(\sigma_{\mathbf{m}}\rho + \rho\sigma_{\mathbf{m}} -2\Braket{\sigma_{\mathbf{m}}}\rho)dW , \;\;\;\;\;   \label{eqqnofb} 
\end{eqnarray}
where $k$, referred to as the \textit{measurement strength}, determines the rate at which the measurement extracts information, and $dW$ is an increment of Wiener noise~\cite{JacobsSP}. The continuous stream of measurement results, $y(t)$, is given by 
\begin{equation}
   dy(t) = \langle \sigma_{\mathbf{m}} \rangle dt + dW/\sqrt{8k} . 
\end{equation}
It is important to note that since the measured observable, $\sigma_{\mathbf{m}}$, can have a $y$ component, it will not leave the density matrix in the form given in Eq.(\ref{rhoxz}). It must therefore be understood that after each time step $dt$, a rotation about the $z$-axis is applied (under which our control problem is invariant) to reduce $a_y$ to zero, and thus keep $\rho$ in the $xz$-plane. 

All Hamiltonians for a single qubit can be parametrized by a single direction around which they rotate the cubit, and a size parameter giving the speed of this rotation. We can write general Hamiltonian as 
\begin{equation}
   H_{\mathbf{n}} = \hbar(\mu/2)\mathbf{n}\cdot\boldsymbol{\sigma} , 
\end{equation}
where $\mathbf{n}$ is the real, unit norm, three-dimensional vector that gives the axis of ration, and $\mu$ gives the angular speed of rotation. We choose as our target state the ground state, $|0\rangle$, for which the Bloch vector is $(0,0,-1)$. Since the density matrix lies in the $xz$-plane, the Hamiltonian that rotates the qubit closest to the target state in a time step $dt$, for a given value of $\mu$, is 
\begin{equation}
 H(t)  = \mbox{sgn}[\theta(t)] \hbar (\mu/2) \sigma_y .  
  \label{fbham1} 
\end{equation}

\textit{Bounds on the control speed}: The natural constraints that we place on the speed of the controls are i) a bound on the speed of the Hamiltonian, and ii) a bound on the rate at which the measurement can extract information. These bounds are 
\begin{equation}
   \mu \leq  \omega ,  \;\;\;\; \mbox{and}  \;\;\;\; k \leq k_{\kms{max}} ,  
\end{equation}
for some positive constants $\omega$ and $k_{\kms{max}}$. We allow the controller to measure in any basis, and apply a Hamiltonian that rotates in any direction. 

\textit{Control objective:} The objective of the control is to maximize the probability, $P$, that the qubit will be found in the target state, $|0\rangle$, in the steady-state. The regime of good control is defined by $\varepsilon \ll 1$, where $\varepsilon \equiv 1 - P$ is the error probability. In this regime we can write the error probability as 
\begin{equation}
    \varepsilon =  1 - (1 + a \cos\theta)/2 = \Delta/2 + a \theta^2 /4 + \mathcal{O}(\theta^4) , 
\end{equation}
were we have defined $\Delta \equiv 1 - a$. In the regime of good control, the qubit spends most of its time in states for which $\theta$ and $\Delta$ are small parameters.


We will assume our qubit is driven by thermal noise, for which the master equation is~\cite{Breuer07} 
\begin{equation}
   \dot{\rho} = \frac{\gamma}{2} (n_T + 1) \mathcal{D}(\sigma)\rho  + \frac{\gamma}{2} n_T \mathcal{D}(\sigma^\dagger)\rho,   \label{thermme}
\end{equation}
where $\mathcal{D}(c) \equiv   2 c \rho c^\dagger - c^\dagger c \rho - \rho c^\dagger c$. Here $\sigma = (\sigma_x - i\sigma_y)/2 = |0\rangle \langle 1|$ is the lowering operator, $\gamma$ is the damping rate, and $n_T$ is determined by the temperature and the energy gap between the ground and excited states of the qubit. The excited-state population at thermal equilibrium is given by $P_T^{\kms{e}} = n_T/(1+2n_T)$. Thermal noise includes decay as a special case ($n_T=0$), and as noted above, the arguments we employ below apply also to dephasing and depolarizing noise. 


\section{Simplifying the optimization problem}
\label{args}

We now present a number of arguments, each of which suggest that we can eliminate one or more parameters from the control problem, while exactly or approximately preserving the optimality of the protocol. Eliminating these parameters greatly simplifies the control problem. 

The first argument is that, since we can measure in any basis and apply any Hamiltonian, it is reasonable to expect that we are always in a better position for the purposes of future control when the Bloch vector is closer to the state $|0\rangle$, given that its length is fixed. While intuitively obvious, this statement has not been rigorously proved, to authors knowledge. Nevertheless this suggests that we should choose the Hamiltonian at each time-step to rotate the state towards the target at the maximum possible speed, since this achieves the closest state to the target at the end of the time-step, over all choices of the Hamiltonian.  This means setting $\mu = \omega$, and choosing the Hamiltonian as given in Eq.(\ref{fbham1}). 

The second argument is that, since we can measure in any basis, it is always best to measure at the maximum strength. The reason for this is that we can always choose to measure in the eigenbasis of the density matrix, a measurement that is effectively classical. As such it reduces the entropy of the system without introducing any quantum back action, and as such does not change the direction of the Bloch vector. Since our goal is to achieve a state that is as pure as possible, this choice of measurement produces a benefit without any detriment to the future control. This argument suggests that we can set $k = k_{\kms{max}}$ and still maintain optimality, leaving us only with the basis of the measurement undecided. 

The third argument concerns the basis of the measurement. The question is whether we can restrict ourselves to measurements that keep the Bloch vector in the $xz$-plane, or whether we should include measurements of observables that include a component of $\sigma_y$. While motion in the $y$ direction is irrelevant for the purposes of our control protocol, measurements with strength $k_{\kms{max}}$ that include a component of $\sigma_y$ will have a different effect on $\theta$ and $a$ than those that do not, and the dynamics of these variables is relevant. We now note that we are most interested in the regime of good control, in which $k_{\kms{max}}$ and $\omega$ are large enough that the feedback protocol can keep $1 - P \ll 1$, and thus $1 - a \ll 1$ and $\theta \ll 1$. For small $\theta$ the difference between a measurement along $\sigma_x$ and one along $\sigma_y$ is second order in $\theta$. Thus in the regime of good control, making a measurement with a component of $\sigma_y$ can only have a minor effect on the performance. We can therefore restrict ourselves to measuring observables of the form 
\begin{equation}
   \sigma_{\mathbf{m}} = \sin(\alpha) \sigma_x - \cos(\alpha) \sigma_z , 
\end{equation}
and so are described by a single angle $\alpha$. If $\alpha=\theta$ then the measurement is ``aligned'' with the state, and thus is in the eigenbasis of the density matrix. In this case the measurement causes no diffusion in $\theta$. 

\begin{figure}
\begin{center}
		\includegraphics[width=0.42\hsize]{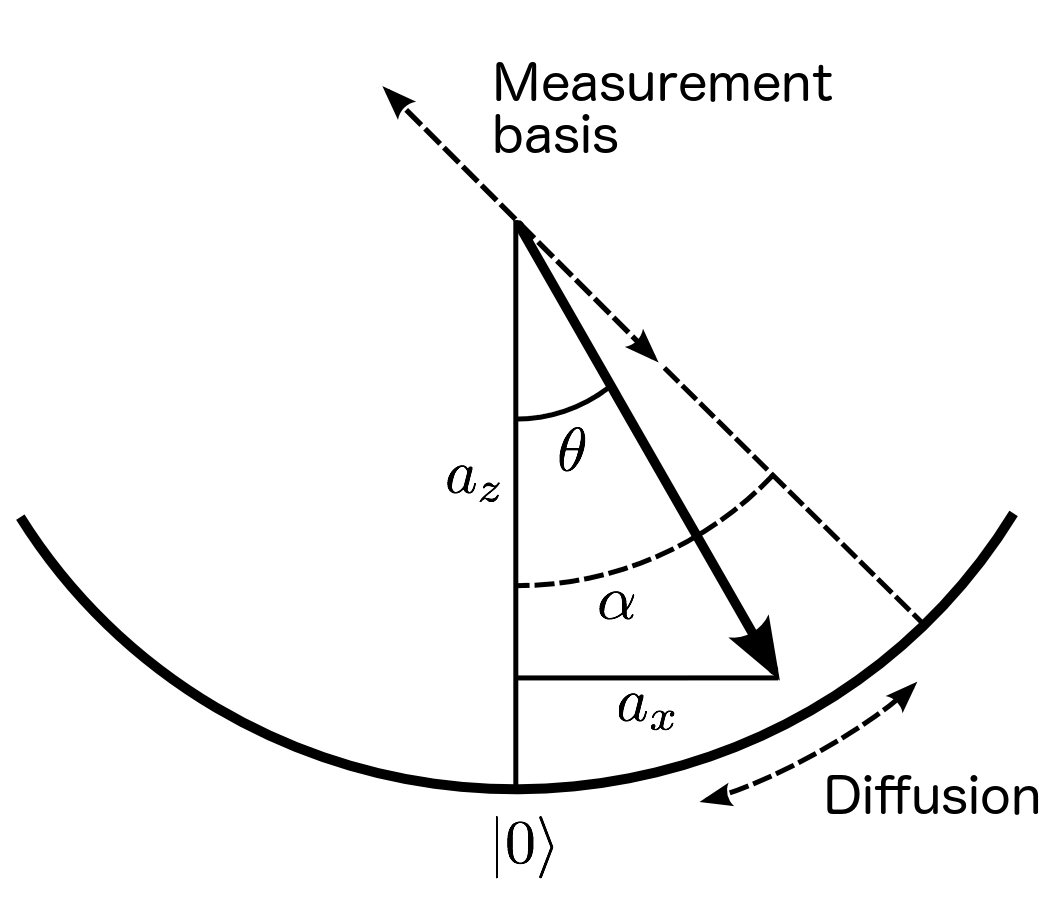}
\caption{Here we depict various elements of the feedback control protocol. The solid arrow is the Block vector, which lies in the $xz$-plane at an angle $\theta$ to the $z$-axis, and the solid curve is the surface of the Bloch sphere. The measurement basis is at an angle $\alpha$ with respect to the $z$-axis, and the ground state $\Ket{0}$ is at the bottom of the sphere. The measurement causes diffusion in $\theta$ if $\alpha \not= \theta$.}
\label{Scheme}
\end{center}
\vspace{-4ex}
\end{figure} 

The forth and final argument is that in the regime of good control the measurement basis, defined by $\alpha$, need not depend upon the length of the Bloch vector, $a$. This insight comes from examining the equations of motion for $a$ and $\theta$ from a measurement at the angle $\alpha$.  The simplest way to derive these equations is to have the Bloch vector point directly upwards, so that $a = a_z$ and $a_x = a_y = 0$. Since we can always chose the Bloch vector as our axis of quantization if we wish, this case provides all the information we need about the dynamics. We write $\rho$ in terms of the Bloch vector, and substitute this in Eq.(\ref{eqqnofb}) to derive the equations of motion for $a_x$ and $a_z$. From these we use Ito's rule to obtain the equations of motion for $a = \sqrt{a_x^2 + a_z^2}$ and $\theta = \tan(a_x/a_z)$, which are 
\begin{eqnarray}  
   d\theta & = &  2k \sin ( 2\alpha) \left(3 -  \frac{2}{a^2} \right) dt + \sqrt{8k} \sin(\alpha) \left( \frac{1}{a} \right) dW  \label{eq::dtheta} \\ 
    da & = &  4k \sin^2(\alpha) \left(a -  \frac{2}{a} \right) dt + \sqrt{8k} \cos(\alpha) (1 - a^2) dW \label{eq::da}  . 
\end{eqnarray} 
When $a$ is close to unity, the regime of good control, we can expand these equations as a power series in $\Delta = 1 - a$. Keeping terms up to first order in $\Delta$ we have  
\begin{eqnarray}
   d\theta & = &  2k \sin ( 2\alpha) \left(1 - 4 \Delta \right) dt + \sqrt{8k} \sin(\alpha) \left( 1 + \Delta \right) dW  \label{eq::dtheta2} \\ 
    da & = &  - 4k \sin^2(\alpha) \left(1 + 3\Delta \right) dt + 2\Delta \sqrt{8k} \cos(\alpha) \, dW \label{eq::da2}  . 
\end{eqnarray} 
We can now easily generalize these equations for $\theta \not=0$ if we wish, which is achieved by the replacement $\alpha \rightarrow \alpha-\theta$. From these equations we see immediately that the leading order terms in the motion of $\theta$ are of order unity, and it is only the next-to-leading-order terms that depend on $\Delta$, since $\Delta \ll 1$. The length of the Bloch vector therefore has little effect on the dynamics, and thus the control, of $\theta$ in the regime of good control. Examining the equation of motion for $a$ we see that to leading order the deterministic part of this equation (the term multiplying $dt$) is also independent of $\Delta$, and thus $a$, but this is not true of the stochastic part (the term multiplying $dW$). The fact that the stochastic part is proportional to $\Delta$ is precisely the diffusion gradient induced by the measurement, and by which the measurement increases the length of the Bloch vector (makes the state more pure). The important fact for our purposes is that as far as this diffusion gradient is concerned, making $\alpha$ dependent on $\Delta$ has the same action as changing the measurement strength. On physical grounds it is apparent that modulating the measurement strength (that is, reducing it below its maximal value), is not useful as it cannot increase the rate at which the measurement purifies the state. A more quantitative argument is as follows. If we choose $\alpha$ as a function of $\Delta$, so that the stochastic term is proportional to a higher power of $\Delta$, then we will reduce the diffusion gradient, thus effectively reducing measurement strength. This argument does not apply for powers of $\Delta$ that are less than unity, but numerical simulations show that if we replace $\Delta$ with $\sqrt{\Delta}$, the rate of purification is reduced~\footnote{KJ performed these simulations while working on~\cite{Jacobs2010}, although they were not reported there.}. 

With the above simplifications the state $\rho$ is defined by two parameters, $a$ and $\theta$, and the feedback protocol is completely specified by a function $\alpha = f(\theta)$ that tells us how to chose the measurement angle based on the location of the Bloch vector. Finding the optimal $f(\theta)$ is a task that is feasible on a parallel computer. We depict the geometry of the control protocol in Fig.~\ref{Scheme}.

\section{Quantum verses classical feedback control}

Once we have reduced a quantum control problem to a set of differential equations that tell us how the controls effect the dynamics of the system (for example, Eqs.(\ref{eq::dtheta}) and (\ref{eq::da})) then the difference between classical and quantum control is merely in what kind of dynamics arises. In classical control is it not usual for the measurement to effect the noise that drives the system, and this is the primary difference. For completeness, if we include the feedback Hamiltonian that always rotates the state towards the target, then the equations that define our control problem (having used the first three of the above simplifications) are 
\begin{eqnarray}  
\hspace{-1.5cm}   d\theta & = & -\mu \, \mbox{sgn}(\theta) dt - 2k \sin ( 2[\alpha-\theta]) \left(3 -  \frac{2}{a^2} \right) dt + \sqrt{8k} \sin(\alpha-\theta) \left( \frac{1}{a} \right) dW  \label{eq::dtheta} \\ 
   \hspace{-1.5cm}  da & = &  4k \sin^2(\alpha-\theta) a \left(1 -  \frac{2}{a^2} \right) dt + \sqrt{8k} \cos(\alpha-\theta) (1 - a^2) dW \label{eq::da}  . 
\end{eqnarray} 
In this control problem the control parameters $\alpha$ and $k$ change the noise driving the system as well as the deterministic motion, and the result is a complex nonlinear problem. 

In fact, the distinction between classical and quantum control contains a further subtlety that is worth elucidating. Since Eqs.(\ref{eq::dtheta}) and (\ref{eq::da}) describe the evolution of the density matrix, which is the observer's full state of knowledge about the system, these equations are the equivalent of classical equations of motion for a probability density in phase space that gives the observer's state of knowledge about the classical system. The Wiener noise driving the equations is therefore not the noise driving the system that we wish to minimize, but noise that tells us the random change in our state of knowledge due to the stream of measurement results. This noise necessarily depends on the measurement, and does so also in the classical equation of motion for the probability density. The real difference between quantum and classical is that, since the classical measurement does not disturb the system, there is a dynamical model underlying the classical probability density in which the system is driven by noise, and controlled by the feedback forces, and this dynamics is not affected by the measurement. This means that the optimization of the measurement and that of the controls can often be separated. While the controls can affect how much information the measurement obtains (because this information may depend on the state of the system) the controls can usually be optimized without reference to the measurement. In the quantum case, if it is even possible to construct an underlying model then the noise driving the system does depend on the measurement. Thus in the control problem given by Eqs.(\ref{eq::dtheta}) and (\ref{eq::da}) we must consider the measurement parameters $\alpha$ and $k$ as a integral part of the optimization of the control protocol.

\section{Numerical optimization}
\label{nums} 

To find the function $\alpha = f(\theta)$ via numerical optimization we must discretize it. The fact that $\theta$ is small suggests that $f(\theta)$ may be well-approximated by the first few terms of a power series. We set $\alpha = \sum_{n=0}^3 c_n \theta^n$ and perform a gradient search using the BFGS quasi-Newton method~\cite{Nocedal06} to find the values of $(c_0,\ldots,c_3)$ that minimize $\varepsilon$. For these simulations we measure time in units of $k$ (that is, we set $k=1$) and use $\gamma = 0.1k$ and $n_T = 0.1$, so that we are in the regime of good control. (We find that good control requires $k \gg \mbox{max}(\gamma, n_T \gamma)$). The initial state of the qubit the thermal state at the ambient temperature, which is the steady-state of the master equation given in Eq.(\ref{thermme}). We run the control protocol for long enough that the qubit settles down to a steady-state under the control, and we are therefore in the regime of good control. Note that the steady-state is given by averaging over the many trajectories, corresponding to the many possible streams of measurement results. In any given trajectory the state of the qubit continues to evolve under the feedback.  

The results of running the optimization for a range of values of the feedback strength, $\omega$, are enlightening. The error $\langle \varepsilon \rangle$ is dominated by the first two parameters in the power series expansion, $c_0$ and $c_1$. Within the statistics of our results, in which we averaged over $128000$ noise realizations, the values of $c_2$ and $c_3$ have no significant effect on the performance. In view of this we simplify the class of protocols in our search space further by keeping only $c_0$ and $c_1$: $\alpha$ is now a linear function of $\theta$. 

\begin{figure}[Bifurcation]
\begin{center}
		\includegraphics[width=0.65\hsize]{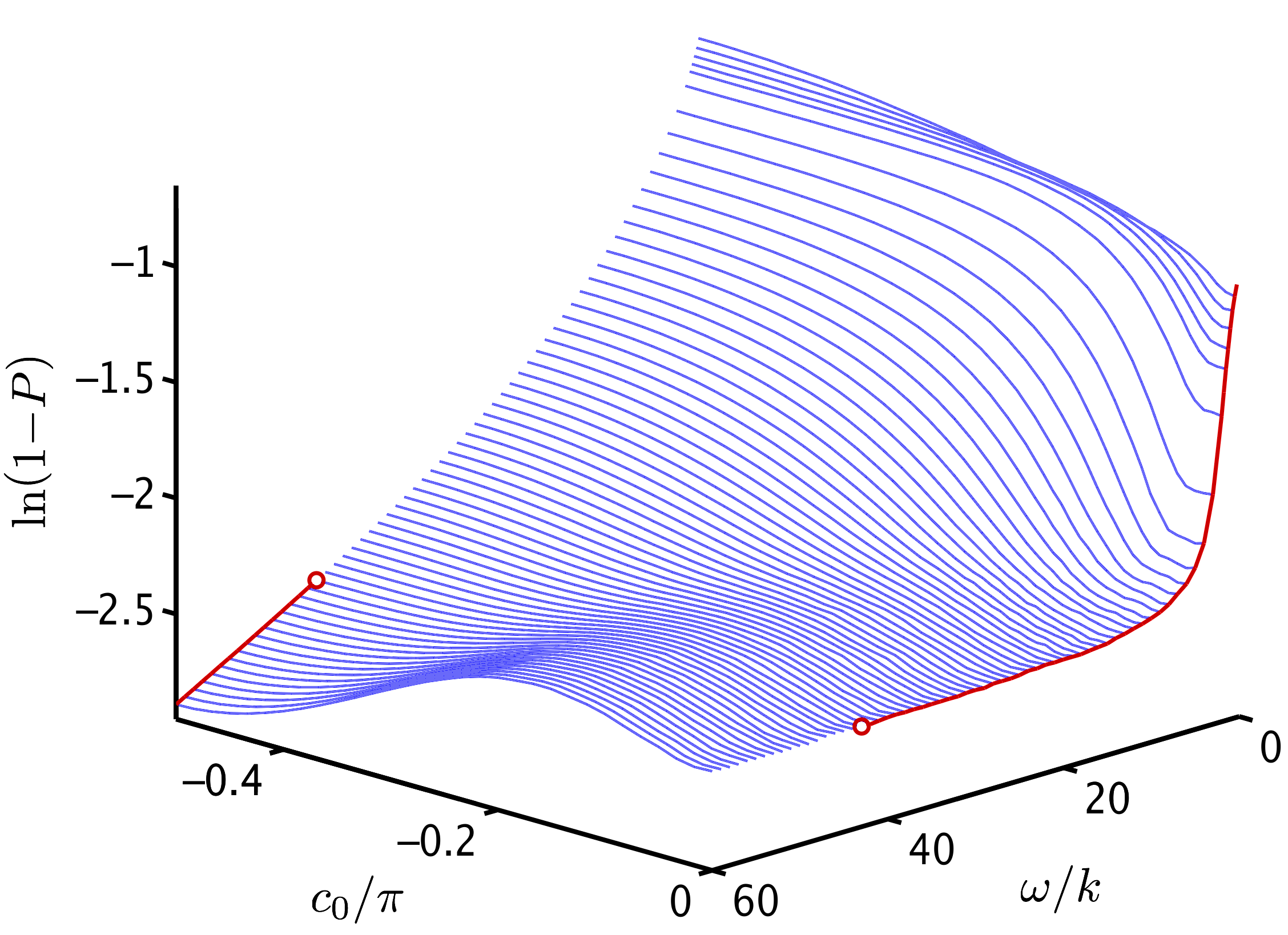}
\caption{Graph of the base-10 logarithm of the steady-state error, $\varepsilon$, vs. the control parameter $c_0$ and feedback strength ($\omega$), with the value of $c_1$ given by Eq.(\ref{eqc1}).  Our protocol is defined by the parameters $c_0$ and $c_1$, and is defined in Eqs.(\ref{fbham}) through (\ref{eqc1}). The dark lines show i) the value of $c_0$ for our protocol as a function of $\omega$, and ii) the performance of our protocol which is the result of optimizing over $c_0$ and $c_1$. The discontinuity in the protocol occurs at $\omega \approx 45k$. We also show the performance as a function of $\omega$ in Fig.~\ref{fig3}.}
\label{optb}
\end{center}
\vspace{-4ex}
\end{figure}

To find the best protocol for each value of $\omega$ we must explore the performance as a function of our three parameters, $c_0$, $c_1$, and $\omega$. Using the same values for $k$, $\gamma$, and $n_T$ as above we calculate $\langle \varepsilon \rangle$ for the full range of values of $c_0$, and for $c_1 \in [-2,2]$, for a discrete set of values of $\omega$. These results, given in the supplementary material, show that the minimum is always at $c_0 = 0$ for $\omega \lesssim 45k$, regardless of the value of $c_1$, and at $|c_0|=\pi/2$ for $\omega \gtrsim 45k$, regardless of the value of $c_1$. At $\omega \approx 45k$ the values $c_0=0$ and $|c_0| = \pi/2$ give the same performance, at least to the accuracy of our results. The fact that the optimal landscape has this structure considerably simplifies the task of finding the optimal values of $c_1$, and thus determining the full control protocol. All we have to do is to find the optimal values of $c_1$ along the two line segments defined by $(c_0=0, 0 < \omega < 45k)$ and $(c_0 = \pi/2, \omega > 45k)$. We find that $c_1$ does not have a significant effect on the performance for $\omega \gtrsim 30k$, and so for the second line segment its value is unimportant. For $c_0 = 0$ we obtain the optimal value of $c_1$ as a function of $\omega$ by hand, and find that the exponential function given in Eq.(\ref{eqc1}) fits the data points quite well, with the parameters $A$, $B$, and $r$ given in table~\ref{tab1} for three values of $\gamma$. In fact, the noise in, and resolution of, our data points means that they have significant fluctuations around this fitted function. Since we do not know that the optimal value of $c_1$ really follows the exponential function, the fluctuations of our data points about the fitted curve are a better measure of the error in our choice of $c_1$ than the estimated errors in the fitted parameters $A$, $B$, and $r$. The mean, $m$, and standard deviation, $\sigma$, of these fluctuations are also given in table~\ref{tab1}. As an example of the significance of $c_1$, for $\gamma = 0.1$ and $\omega = 10k$, choosing the optimal value of $c_1$ ($\sim 0.7$) gives a steady-state error of $\varepsilon = 3.3\times 10^{-3}$, whereas setting $c_1=0$ gives $\varepsilon = 4.6\times 10^{-3}$. A change in $c_1$ of $0.01$ (the level of our uncertainty in the optimal value) changes $\varepsilon$ by less than $5\%$.  As $\omega$ increases the importance of $c_1$ decreases: for $\omega = 20k$, setting $c_1 = 0.7$ gives $\varepsilon = 3.0\times 10^{-3}$, and $c_1 = 0$ gives $\varepsilon = 3.4\times 10^{-3}$.  

\begin{figure}
\begin{center}
		\includegraphics[width=0.5\hsize]{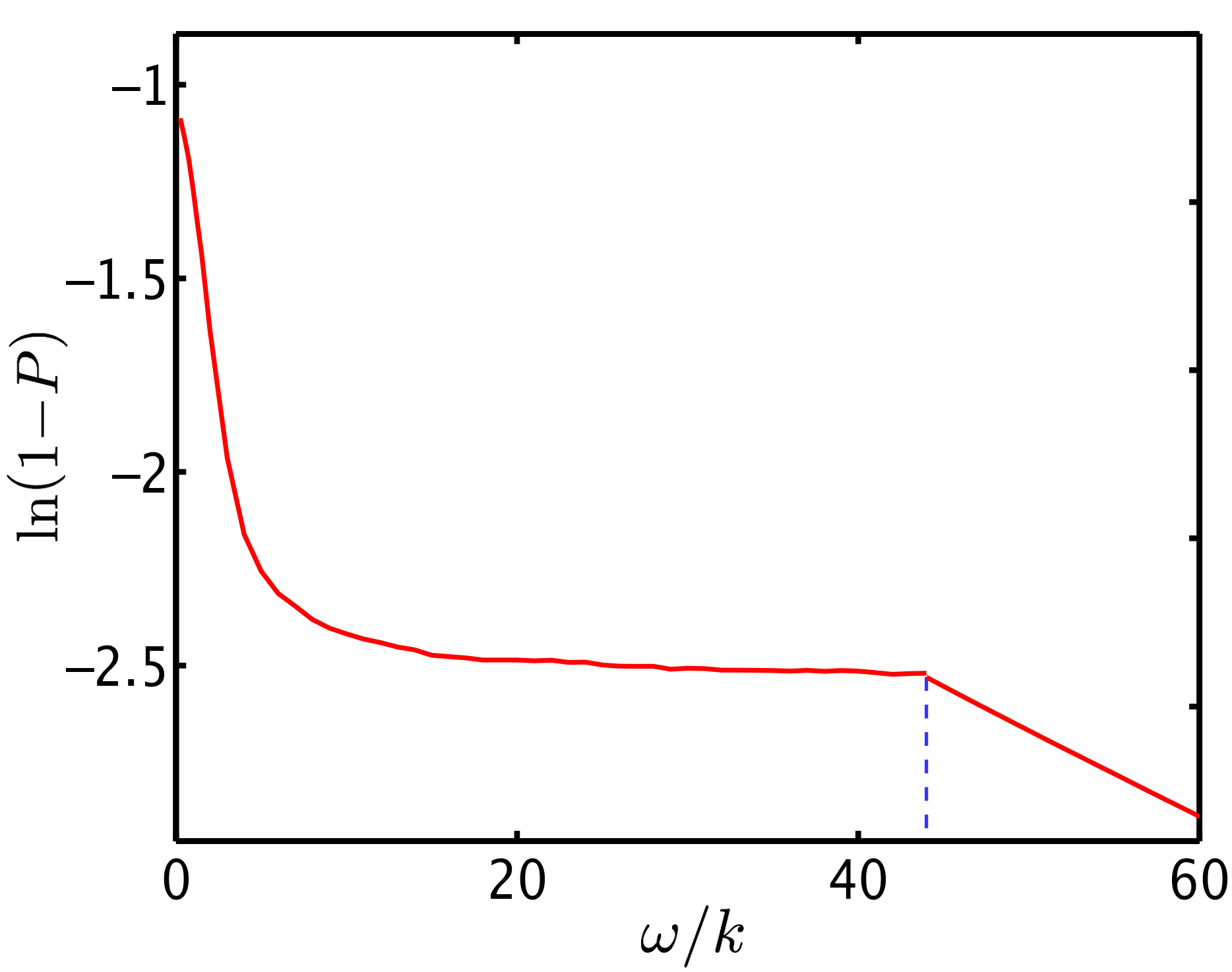}
\caption{Graph of the performance of our protocol as a function of $\omega$. This plot corresponds to the dark lines in the plot in Fig.~\ref{optb}. The dashed line is at $\omega=44k$, and marks the approximate location of the discontinuity, at which point the optimal value of $c_0$ switches from $0$ to $\pi/2$.}
\label{fig3}
\end{center}
\vspace{-4ex}
\end{figure}

\begin{figure}
\begin{center}
		\includegraphics[width=1\hsize]{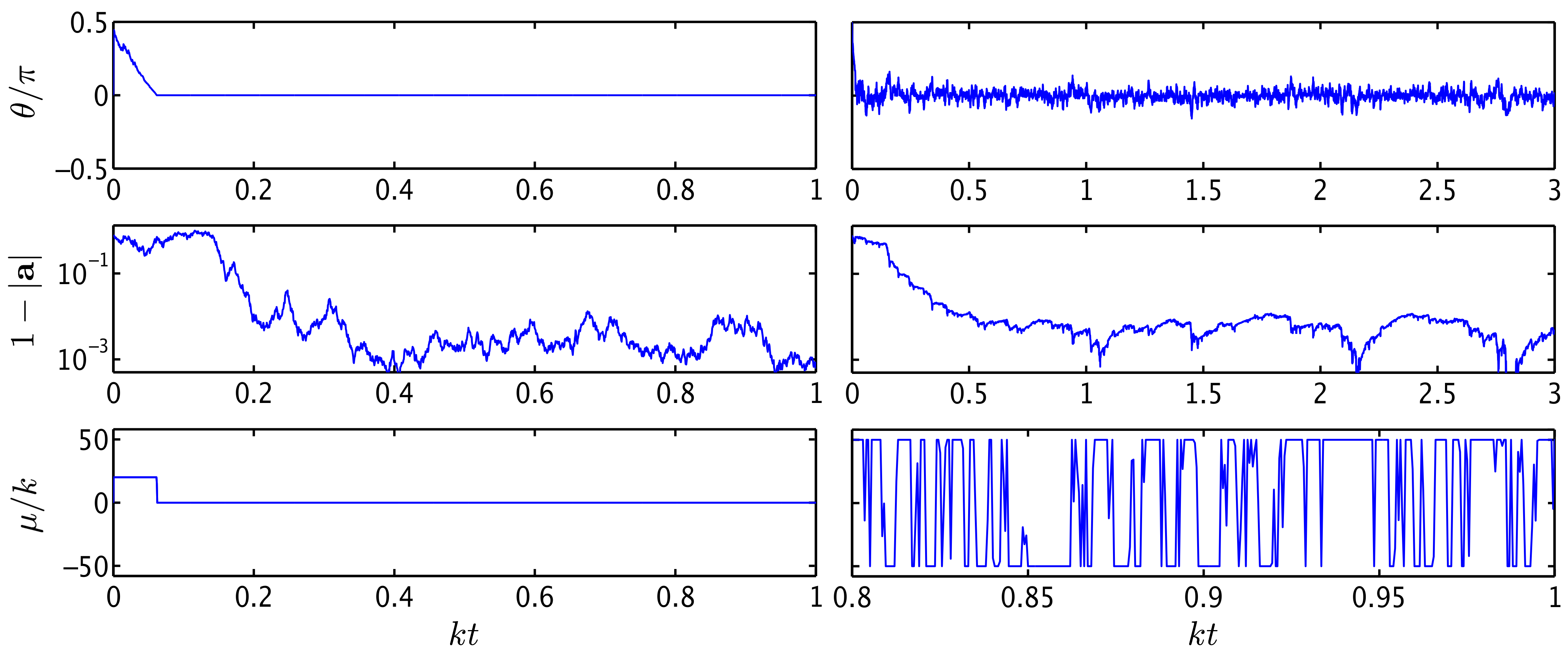}
\caption{Here we show two sample trajectories for a qubit under our feedback protocol. We chose the initial state $\mathbf{a} = (1/2,0,0)$. On the left is a trajectory for $\omega = 20 k$ and on the right for $\omega = 50 k$. Note that the plot for the feedback Hamiltonian for the latter (bottom right) is a blowup on the time axis so that the rapid fluctuations of the feedback rotation speed, $\mu$, are visible.}
\label{fig4} 
\end{center}
\vspace{-4ex}
\end{figure}

\section{The protocol}

We can now summarize the results in the previous section as the following feedback protocol. The feedback Hamiltonian is chosen to be
\begin{equation}
 H(t)  = \mbox{sgn}[\theta(t)] \hbar (\omega/2) \sigma_y .  
  \label{fbham}
\end{equation}
The measurement is made at the maximum rate $k_{\kms{max}}$, and the measured observable is chosen to be 
\begin{equation}
  \sigma_{\alpha} = \sin\alpha \, \sigma_x - \cos\alpha \, \sigma_z  
\end{equation}
with  
\begin{equation}
 \alpha = c_0 + c_1\theta(t) , 
\end{equation}
and  
\begin{eqnarray} 
 c_0  & = &  \left\{  \begin{array}{cc}
                                 0  &  \omega \lesssim 45 k   \\
                                \pi/2  &   \omega \gtrsim 45 k   
                                \end{array}  \right.  ,   \\
 c_1  & = &  - A - B [ 1 - e^{-r \omega /k} ]  . \label{eqc1}
\end{eqnarray}


To show how the optimal performance, defined as the minimum steady-state error, depends on $\omega$ we now plot the performance as a function of $c_0$ and $\omega$ in Fig.~\ref{optb}. In this plot we set the value of $c_1$ to that given by Eq.(\ref{eqc1}). This choice gives the best performance (the performance of our protocol) for $c_0 = 0$ and for  $|c_0| = \pi/2$ for $\omega > 45k$ (that is, when $c_0$ has its optimal value), since in the latter case  the value of $c_1$ is unimportant. So the plot gives the performance of our protocol, but does not show the best performance that can be obtained when $c_0$ is outside its optimal value and $\omega \lesssim 30k$. In Fig.~\ref{fig3} we again show the performance of our protocol (given by the dark lines in Fig.~\ref{optb}), but this time only as a function of $\omega$ so that the performance can be read-off more easily.   

We display sample trajectories in Fig.~\ref{fig4} for the two kinds of protocols, that for $k > 45 \omega$ (weak feedback) and $k < 45\omega$ (strong feedback). In the former, displayed on the left, the measurement is aligned with the state as $\theta \rightarrow 0$, with result that fluctuations in $\theta$ are virtually zero in the steady-state (at least for the value for $\omega$ that we use in the figure). Because of this the Hamiltonian is only required for the initial transient to bring $\theta$ to $0$. For strong feedback the measurement generates continual diffusion for $\theta$, and the Hamiltonian switches continually to combat this diffusion. Note that while the protocol specifies that the feedback rotation speed $\mu$ should switch between $\pm \omega_{\mss{max}}$, for a numerical simulation with finite step-size $\Delta t$, (and in applications) $\mu$ should be chosen so as not to over-rotate the Bloch-vector in any given time-step. It is for this reason that $\mu$ is not always at its maximal value.  

\section{Discussion}
\label{disc}

We have obtained a feedback protocol that can be neatly specified. But we also want to understand why the protocol should have the form it does. It turns out that we can understand the main features of the protocol in terms of three known dynamical effects of continuous measurement. The first is that a measurement in a basis close to that of the Bloch vector tends to ``drag'' the Bloch-vector in the direction of the measurement. This effect is often referred to as the ``quantum anti-Zeno'' effect~\cite{Balachandran00}, and it explains why the co-efficient $c_1$ is negative: this causes the measurement to drag the state towards $|0\rangle$, and thus makes the most use of it. The second effect comes from the fact that measuring at an angle $\alpha \not= \theta$ generates diffusion for $\theta$. The amount of diffusion is proportional to $\sin(|\theta-\alpha|)$, and a gradient in the diffusion rate pushes the state into regions of low diffusion~\cite{Jacobs2010}. Our protocol states that when there is no feedback Hamiltonian ($\omega = 0$) we should should set $\alpha = -\theta/2$. This means increasing the difference between $\alpha$ and $\theta$, away from the target state, thus increasing the diffusion. The resulting diffusion gradient  pushes the state towards $\theta=0$.  

\begin{table}[t]
\caption{Values for the parameters of Eq.(\ref{eqc1})}
\begin{center}
\begin{tabular}{cccccc}
$\gamma$ & $A$ & $B$ & $r$ & $m$ & $\sigma$ \\
\hline 0.1  $\;$  & 0.500 $\;$ & 0.186 $\;$ &  0.476 $\;$ & 0.002 $\;$ & 0.007  \\
           0.2 $\;$ & 0.479 $\;$ & 0.211 $\;$ & 0.705 $\;$ & -0.005 $\;$ & 0.011 \\
           0.3 $\;$  & 0.478 $\;$ & 0.217 $\;$ & 0.529 $\;$ & 0.001 $\;$ & 0.008 
\end{tabular}
\end{center}
\vspace{-5ex}
\label{tab1}
\end{table}%

We note that it is possible to derive the optimal value of $c_1$ for $\omega = 0$ from an approximate calculation. Assuming that the system stays close to the target state throughout its evolution so that we can set $\cos\theta \approx 1 - \theta^2/2$, and setting $a \approx 1$, we obtain the following equation of motion for $\theta^2$ under the measurement and feedback:
\begin{eqnarray}
  d(\theta^{2}) & = & \{4\gamma n_T - 2\omega|\theta|+[8kc_1(c_1+1) -\gamma]\theta^{2} \} dt \nonumber \\
 & & +\, \sqrt{2k}(1+c_1)\theta^{2}dW . 
  \label{analytic}
\end{eqnarray}
If we set $\omega = 0$ (no feedback Hamiltonian) and take the average on both sides, then we obtain a stochastic equation for $\langle \theta^2 \rangle$ that can be solved analytically~\cite{JacobsSP}. Solving this equation for the steady-state shows that the minimum value of $\langle \theta^2 \rangle$ occurs at $c_1 = -0.5$.  

The third effect is important in the regime of strong feedback ($\omega > 45k$). In this regime our protocol tells us to measure approximately at right angles to the Bloch vector, causing the maximum diffusion in $\theta$. This can be understood from the following property of measurement: the average rate at which the measurement purifies the state (that is, lengthens the Bloch vector), is greatest when the diffusion is greatest. When the feedback Hamiltonian is sufficiently fast ($\omega \gg k$) it can suppress the unwanted diffusion and thus take advantage of the increased purification. That this would be true for sufficiently strong feedback was already known~\cite{rapidP, Shabani08} ---  what is unexpected is that the optimal value of $c_0$ switches abruptly from $0$ to $\pi/2$ at a given value of $\omega/k$. 


To summarize, we have obtained a feedback control protocol for a single qubit that gives a nontrivial prescription for choosing the measurement angle as a function of the direction of the Bloch vector and the feedback strength. We conjecture that this protocol is optimal (to first-order in $\theta$) in the regime of good control. Time will hopefully tell if this conjecture is correct. 

\textit{Acknowledgments}: This work was performed with the use of the supercomputing facilities managed by the Research Computing Department at the University of Massachusetts Boston. KJ is supported by the NSF projects PHY-1005571 and PHY-1212413, and by the ARO MURI grant W911NF-11-1-0268. 

\vspace{5mm}


\end{document}